\documentclass[aps,prb,preprint]{revtex4}
\usepackage{amssymb,latexsym,amsmath}
\usepackage{graphicx}
\usepackage{color}
\usepackage{subfig}
\usepackage{hyperref}

\begin{document}

\title{The Rigorous Stochastic Matrix Multiplication Scheme for the Calculations of Reduced Equilibrium Density Matrices of Open Multilevel Quantum Systems}
\author{Xin Chen}
\affiliation{Department of Chemistry,  \\ Massachusetts Institute of Technology, \\ Cambridge, MA 02139}
\date{now}
\begin{abstract}
Understanding the roles of the temporary and spatial structures of quantum functional noise in open multilevel quantum molecular systems attracts a lot of theoretical interests. I want to  establish a rigorous and general framework for functional quantum noises from the constructive and computational perspectives, $\it{i.e.}$ how to generate the random trajectories to reproduce the kernel and path ordering of the influence functional with effective Monte Carlo methods for arbitrary spectral densities. This construction approach aims to unify the existing stochastic models to rigorously describe the temporary and spatial structure of Gaussian quantum noises. In this paper, I review the Euclidean imaginary time influence functional and  propose the stochastic matrix multiplication scheme to calculate reduced equilibrium density matrices (REDM).

In addition, I review and discuss the Feynman-Vernom influence functional according to the Gaussian quadratic integral, particularly its imaginary part which is critical to the rigorous description of the quantum detailed balance.
As a result, I establish the conditions under which the influence functional can be interpreted as the average of exponential functional operator over real-valued Gaussian processes for open multilevel quantum systems. I also show the difference between the local and nonlocal phonons within this framework.
With the stochastic matrix multiplication scheme, I compare the normalized REDM with the Boltzmann equilibrium distribution for open multilevel quantum systems.

\end{abstract}
\maketitle
\newpage

\section{Introduction}
 Rigorously calculating the time evolution and thermal properties of open quantum systems is a long-time challenging question. Many different  second-order perturbative master equations\cite{fleming,kampen} have been extensively used to study varieties of transport processes, which are valid for the weak coupling limit. Subsequently, having a rigorous computational method to calculate the equation of motion and steady state of reduced quantum systems is a challenging task, particularly for the intermediate system-bath coupling. The hierarchical equation of motions developed originally by Kubo and his coworkers \cite{kubo} assuming exponential kernels, $\it{i.e.}$ Gauss-Markov model at the high temperature, is still not feasible for large quantum systems and general temporary and spacial structures of quantum noise.

 In order to achieve the goal,  novel and rigorous conditional path integral Monte Carlo method was proposed \cite{chencaosilbey} and developed  to account for the complex fluctuating environment trajectories and calculate the reduced density matrix dynamics and steady state according to the (Euclidean/Feynman-Vernon) influence functional formalism. In this development, due to the quantum path interferences in multi-level open quantum systems, it is necessary to consider the hidden structures of Gaussian processes. As a result, the major obstacle is how to sample quantum noise with a rigorous stochastic interpretation of the Euclidean/Feynman-Vernon influence functional with temporary and spatial (or different energy surfaces) structures for multilevel quantum systems. The previous work shows that quantum Gaussian functional noise\cite{cnumber} is governed by the second order time correlation functions (kernel).  For some special limits, such as Gauss-Markovian\cite{kubo}, semi-classic limit\cite{cnumber,gao}, Gaussian white noise\cite{haken},  $\it{etc}.$, explicit propagation schemes have been proposed. However, no rigorous and general theory of how to construct a Gaussian process for arbitrary spectral densities is established.

This paper is devoted to the development of the stochastic matrix multiplication scheme to calculate REDM for open multi-level quantum systems, which is related to the idea of matrix multiplication scheme proposed before\cite{berne} to calculate canonical  density  matrix. Particularly, I aim to establish the conditions under which both the Feynman-Vernon (real time) and Euclidean (imaginary time) influence functional can be interpreted as Gaussian moment-generating/characteristic functions. The future directions include the extensions to rigorously address full rank coupling matrix, quantum path interference and real time dynamics using techniques such as analytic continuation\cite{berne1}, remapping of the complex-valued Gaussian processes \cite{chencaosilbey} to real-valued ones, effective sampling strategies\cite{liu} and $\it{etc.}$.

Non-equilibrium steady states \cite{nitzan,nitzan2,jeremy,zhu} are important quantities to model the macroscopic transport processes in open systems. The reduced density matrix of an open quantum system in the long time steady state is approximately equal to REDM to the second order\cite{redm1,Hu}, particularly its off-diagonal terms which is important for the quantum path interference.
As a result, REDM will be a more effective alternative quantity to calculate to describe the macroscopic transport process. With the proposed stochastic matrix multiplication scheme for multi-level quantum systems, REDM can be calculated in a rigorous way. The details of the constructive approach of the stochastic matrix multiplication scheme will be presented by introducing and comparing with the discrete and continuous Gaussian moment generating functions.
Due to the second order approximate relation between the non-equilibrium steady state and REDM, the potential  extensions and applications of the stochastic matrix multiplication scheme in classical force-fields with equilibrium quantum distributions\cite{alan}, quantum heat engine\cite{kosloff,heateng}, non-adiabatic quantum dynamics\cite{coalson}, $\it{etc.}$ can be fruitful and insightful. In the future work, I want to construct and extend Gaussian processes to study the quantum path interference effect due to the incoherent radiative and non-radiative transitions\cite{coalson,scully2}. Particularly, the extension of the imaginary time Gaussian processes to the real time ones to study the reduced density dynamics with the analytical continuation and other technique will be explored as well.

This paper consists of five sections:
1. In Section~\ref{EIF}, I review the Euclidean influence functional; 2. In Section~\ref{genfun}, I discuss the real-valued Gaussian process, its moment generating function and path ordering. Therein, I discuss
the connection between the influence functional, stochastic evolution operator and Gaussian stochastic processes; 3. In Section~\ref{coupling}, I discuss the coupling with local and nonlocal phonons and its implication for the structure of Gaussian processes; 4. In Section~\ref{citpi}, I introduce the stochastic matrix product method and the calculation of REDM; 5. In Section~\ref{results}, I present the results for the simplest dimer systems (spin-boson models) and the calculations of the matrix elements of REDM for different temperatures and parameters of spectral densities. Finally Section~\ref{conrem} is concluding remarks.

\section{System Bath Hamiltonian and Influence Functional}\label{EIF}
The Hamiltonian of an open quantum system with a bilinear coupling to a harmonic oscillator reservoir (phonon) can be defined as,
\begin{equation}\label{Ham}
\hat{H}=\hat{H}_S(q)+\hat{H}_I(q,x_1,x_2,\cdots,x_n)+\hat{H}_B(x_1,x_2,\cdots,x_n),
\end{equation}
where $\hat{H}_I$ is the bilinear coupling $\hat{V}(q)*\hat{\mathbf{X}}(x_1,x_2,\cdots,x_n)$ where $\hat{V}(q)$ is the functional of the system operator $q$, $\hat{\mathbf{X}}=\sum_i g_i x_i$ and $\hat{H}_B = \frac{1}{2} \sum_{i=1}^n m_i (\dot{x}^2_i + \omega_i^2 x_i^2)$. Here, we consider the nonlocal phonon coupling. Detailed discussion can be found in Section~\ref{coupling}.

For the multilevel quantum systems of electronic excitations,
\begin{equation}
\hat{H}_S=\sum_i \epsilon_i \vert i \rangle \langle i \vert + \sum_i \sum_{j\neq i} J_{ij} \vert i \rangle \langle j \vert,
\end{equation}
and its corresponding coupling term,
\begin{equation}
\hat{V} = \sum_i \sum_j V_{ij} \vert i \rangle \langle j \vert.	
\end{equation}

Quantum noise, as the collective motion of  quantum harmonic modes, is different from its corresponding classic version due to the temporary structure of influence functional in REDM,
\begin{equation}\label{einflu}
\rho_{\beta} (q'',q')= \frac {1}{Z} \int_{q(0)=q'}^{q(\hbar \beta)=q''} D q(\cdot) \exp \bigg \{ -S_q^{E} [q(\cdot)] /\hbar \bigg\} \mathcal{F}^{(E)}[q(\cdot)],
\end{equation}
where the influence functional is expressed as,
\begin{equation}
\mathcal{F}^{(E)}[q(\cdot)] \equiv \exp \bigg \{ - S_{infl}^{(E)} [q(\cdot)] /\hbar  \bigg \} = \frac{1}{Z_B} \oint  D \mathbf{x}(\cdot) \exp \bigg \{ - S_{B,I}^{(E)}
[q(\cdot),\mathbf{x}(\cdot)]/\hbar \bigg \}
\end{equation}
with
the effective action,
$S_{B,I}^{(E)}[q(\cdot),\mathbf{x}(\cdot)]=S_B^{(E)}[\mathbf{x}(\cdot)] + S_I^{(E)}[q(\cdot),\mathbf{x}(\cdot)] $.
After reducing the degrees of freedom of the harmonic reservoir, $\mathbf{x}=[x_1,x_2,\cdots,x_n]$,  the effective action in the (Euclidean) imaginary time influence functional \cite{weiss} will become,
\begin{eqnarray}
S_{infl}^{(E)} & = & \int_0^{\hbar \beta} d\tau \int_0^{\tau} d\sigma k(\tau-\sigma) V[q(\tau)] V[q(\sigma)] \\ \nonumber
&& = \frac{1}{2} \int_0^{\hbar \beta} d\tau \int_0^{\tau} d\sigma K(\tau-\sigma) ( V[q(\tau)] - V[q(\sigma)] )^2,
\end{eqnarray}
where
\begin{equation}\label{kernel}
K(\tau)=\frac{1}{\pi} \int_0^{\infty} d \omega J(\omega) D_{\omega} (\tau),
\end{equation}
where $D_{\omega} (\tau) =\frac{\cosh[\omega (\hbar \beta /2 - \tau)]}{\sinh(\omega\hbar \beta/2)} = [1+n(\omega)]e^{-\omega \tau} + n(\omega)e^{\omega \tau}$, and $K(\tau)=\mu:\sigma(\tau):- k(\tau)$ in which $:\sigma(\tau):$ is a periodic delta function and $\mu = \frac{2}{\pi} \int_0^{\infty} d\omega \frac{J(\omega)}{\omega}$.
Naturally, the window from $0$  to $\beta \hbar$ will be used,
\begin{equation}\label{einfl}
S_{infl}^{(E)}  =  \int_0^{\hbar \beta} \mu V[q(\tau)] ^2 - \int_0^{\hbar \beta} d\tau \int_0^{\tau} d\sigma K(\tau-\sigma) V[q(\tau)] V[q(\sigma)].
\end{equation}
This expression is similar to the real-time Feynman-Vernon influence functional which has the forward and backward trajectories and complex kernels,
\begin{eqnarray} \label{infl0}
& &\exp\{-S_{infl}[x(t),x'(t)]\}=\\ \nonumber
& &\exp\bigg \{-\int_{0}^{t}d\tau\int_{0}^{\tau}d\sigma \big [V[q(\tau)]-V[q'(\tau)] \big ] \big \{ \gamma_r(\tau-\sigma) \big [V[q(\sigma)] \\ \nonumber
&& - V[q'(\sigma)] \big ] -i \gamma_i(\tau-\sigma) \big [ V[q(\tau)]+V[q'(\tau)] \big ] \big \} \\ \nonumber
&& + i \frac{\mu}{2} \int_0^t d\tau \big [ V[q(\tau)]^2 -V[q'(\tau)]^2 \big ] \bigg \},
\end{eqnarray}
where $\gamma_r(t)=\textbf{Re}(K(i\,z))=\frac{1}{\pi} \int_0^{\infty} d \omega J(\omega) \coth (\beta \hbar \omega/2) \cos (\omega t)$ and $\gamma_i(t)=\textbf{Im}(K(i\,z)) = - \frac{1}{\pi} \int_0^{\infty} d \omega J(\omega) \sin (\omega t)$ in which $z=t-i\,\tau$\cite{weiss} in which $t$ is the real-time axis and $\tau$ is the imaginary time one. In other words, the real time kernel $K(t-i\,\tau)$ is the analytical continuation of the imaginary-time one $K(\tau)$.

The Gaussian stochastic interpretation of the influence functional is a trick business, particularly for the real time Feynman-Vermon influence functional and  the multi-level quantum systems. The real-valued classic Gaussian process has a symmetric covariance matrix. However, the influence functional doesn't guarantee the symmetry due to the path-ordering. Therefore, in the following sections, I will review the real-time and imaginary-time influence functional and discuss under what conditions the stochastic interpretation of the influence functional can be recovered. At the same time I will construct the Gaussian process to deconvolute the Euclidean influence functional and the Monte Carlo method, stochastic matrix multiplication scheme, to calculate the equilibrium reduced density matrix.

\section{Gaussian Quadratic Integral, Path Ordering and Influence Functional}\label{genfun}

The original paper of the Feynman-Vernon Influence functional\cite{Feynman} has the clear and inspiring discussion on the structures of dynamic quantum Gaussian noise. In Eq. 4.24 and following discussion in the original paper, Feynman and Vernon address the tricky business associated with ``the three
forms of exponents'' in the influence functional. I want to review the discussion from a constructive perspective on the quantum Gaussian noise with comparison to the classic Gaussian noise.
Considering the quadratic Feynman-Vernon influence functional (assuming $\hbar=1$) \cite{weiss},
\begin{eqnarray} \label{infl}
& &\exp\{-S_{infl}[q(t),q'(t)]\}=\\ \nonumber
& &\exp\bigg \{-\int_{0}^{t}d\tau\int_{0}^{\tau}d\sigma V_{-}(\tau)\gamma_r(\tau-\sigma) V_{-}(\sigma)
+ i \int_{0}^{t}d\tau\int_{0}^{\tau}d\sigma V_{-}(\tau) \gamma_i(\tau-\sigma) V_{+}(\sigma) \bigg  \},
\end{eqnarray}
where $V_{+}(\tau)=V[q(\tau)]+V[q'(\tau)]$ and $V_{-}(\tau)=V[q(\tau)]-V[q'(\tau)]$ and $i \frac{\mu}{2} \int_0^t d\tau   V[q(\tau)]^2 -V[q'(\tau)]^2 $ is dropped.
The path ordering $\int_{0}^{t}d\tau\int_{0}^{\tau}d\sigma$ of the influence functional determines that the quantum noise is different from classic noise since
 $\int_{0}^{t}d\tau\int_{0}^{\tau}d\sigma$ is not guaranteed to be equal to $\frac{1}{2} \int_{0}^{t}d\tau\int_{0}^{t}d\sigma$ except for some special cases. This is an important property to quantum noises in existence of quantum path interferences/coherences.
In order to understand the difference between quantum noise and classic noise, it is worthwhile explaining the following subjects including multivariate Gaussian moment generating function (discrete time and continuous time).

\subsection{Discrete Gaussian Process Moment Generating and Characteristic Functions}
The discrete Gaussian process, $\boldsymbol{\hat{\xi}}=[\xi_0,\xi_1,\cdots,\xi_{N}]^T$ of $\xi(t)$ where $\xi_i=\xi(t_i)$ on the homogeneous discrete time grid  $t_0,t_1,t_2\cdots,t_N$ and $t_0=0$ and $t_N=t$, is simply governed by the multivariate Gaussian probability density function (PDF),
\[
\rho (\hat{\boldsymbol{\xi}} ) = \frac{1}{\sqrt{(2\pi)^{N+1} \vert \Sigma \vert }} \exp \bigg(- \frac{1}{2} \boldsymbol{\hat{\xi}}^T \Sigma^{-1} \boldsymbol{\hat{\xi}} \bigg) .
\]
The corresponding moment generating function can be defined as,
\begin{eqnarray}\label{mgf}
& & \frac{1}{\sqrt{(2\pi)^{N+1} \vert \Sigma \vert }} \int_{-\infty}^{\infty} d \boldsymbol{\hat{Q}} \exp(- \boldsymbol{\hat{\xi}}^T \boldsymbol{\hat{Q}} \; dt ) \exp \bigg(- \frac{1}{2} \boldsymbol{\hat{\xi}}^T \Sigma^{-1} \boldsymbol{\hat{\xi}} \bigg) \\ \nonumber
& = & \frac{1}{\sqrt{(2\pi)^{N+1} \vert \Sigma \vert }}  \int_{-\infty}^{\infty} d \boldsymbol{\hat{Q}} \exp( \boldsymbol{\hat{\xi}}^T \boldsymbol{\hat{Q}} \; dt) \exp \bigg(- \frac{1}{2} \boldsymbol{\hat{\xi}}^T \Sigma^{-1} \boldsymbol{\hat{\xi}} \bigg) \\ \nonumber
& = & \exp \bigg(  \frac{1}{2} \boldsymbol{\hat{Q}}^T \Sigma \boldsymbol{\hat{Q}}  \; dt^2 \bigg ),
\end{eqnarray}
where  $\boldsymbol{\hat{Q}}$  is a deterministic discrete process $\boldsymbol{\hat{Q}}=[Q_0,Q_1,Q_2,\cdots,Q_N]^T$ in which $Q_i=Q(t_i)$, $dt=t_i-t_{i-1}$, $\Sigma$ is the covariance matrix of the multivariate Gaussian PDF, $\Sigma = \bigg [ \langle \xi_i \xi_j \rangle \bigg ]$ and $\langle \xi_i \xi_j \rangle = \gamma (t_i-t_j)$ is the two-time auto-correlation function of $\xi(t_i)$ and $\xi(t_j)$ . It is easy to show that the continuous limit, $dt\rightarrow 0$, of the discrete Gaussian moment generating function will become an integral, $\exp (-\frac{1}{2} \boldsymbol{\hat{Q}}^T \Sigma \boldsymbol{\hat{Q}} dt^2 ) \rightarrow \exp (\frac{1}{2} \int_0^t d\tau \int_0^t d\sigma Q(\tau) \gamma(\tau-\sigma) Q(\sigma) )$ due to the even symmetry of $\Sigma$, $\it{i.e.}$, $\gamma (t_i-t_j) = \gamma(t_j-t_i)$.
Therefore, the continuous-time Gaussian moment generating function can be expressed as,
\begin{equation}
\bigg \langle \exp[ \xi(t) Q(t)] \bigg \rangle_{\xi(t)} = \bigg \langle \exp[- \xi(t) Q(t)] \bigg\rangle_{\xi(t)}= \exp \left [\frac{1}{2} \int_0^t d\tau \int_0^t d\sigma Q(\tau) \gamma(\tau-\sigma) Q(\sigma) \right],
\end{equation}
Correspondingly, the characteristic function for discrete and continuous Gaussian processes can be defined as,
\begin{eqnarray*}
& & \frac{1}{\sqrt{(2\pi)^{N+1} \vert \Sigma \vert }} \int_{-\infty}^{\infty} d \boldsymbol{\hat{Q}} \exp(- i \boldsymbol{\hat{\xi}}^T \boldsymbol{\hat{Q}} \; dt ) \exp \bigg(- \frac{1}{2} \boldsymbol{\hat{\xi}}^T \Sigma^{-1} \boldsymbol{\hat{\xi}} \bigg) \\
& = & \frac{1}{\sqrt{(2\pi)^{N+1} \vert \Sigma \vert }}  \int_{-\infty}^{\infty} d \boldsymbol{\hat{Q}} \exp( i \boldsymbol{\hat{\xi}}^T \boldsymbol{\hat{Q}} \; dt) \exp \bigg(- \frac{1}{2} \boldsymbol{\hat{\xi}}^T \Sigma^{-1} \boldsymbol{\hat{\xi}} \bigg) \\
& = & \exp \bigg( - \frac{1}{2} \boldsymbol{\hat{Q}}^T \Sigma \boldsymbol{\hat{Q}}  \; dt^2 \bigg ),
\end{eqnarray*}
and
\begin{equation}
\bigg \langle \exp[ - i  \xi(t) Q(t)] \bigg \rangle_{\xi(t)} = \bigg \langle \exp[ i  \xi(t) Q(t)] \bigg \rangle_{\xi(t)} = \exp \left [-\frac{1}{2} \int_0^t d\tau \int_0^t d\sigma Q(\tau) \gamma(\tau-\sigma) Q(\sigma) \right].
\end{equation}
This is an important equality for the real-time quantum noise.

\subsection{Path Ordering and Gaussian Quadratic Integral}
The path ordering of influence functional requires the following equality to match the Gaussian generating function,
\[
\exp \left [-\frac{1}{2} \int_0^t d\tau \int_0^t d\sigma V(\tau) \gamma(\tau-\sigma) V(\sigma) \right] =
\exp \left [-\int_0^t d\tau \int_0^{\tau} d\sigma V(\tau) \gamma(\tau-\sigma) V(\sigma) \right].
\]
To satisfy the equality, two conditions are needed: the associativity,
\begin{equation}\label{cond1}
V(\tau)V(\sigma) = V(\sigma) V(\tau),
\end{equation}
and the even symmetry of time,
\begin{equation}\label{cond2}
\gamma(\tau-\sigma)=\gamma(\sigma-\tau),
\end{equation}

It is worth emphasizing that in the real-time Feynman-Vernon influence functional of Eq.~\ref{infl}, these two conditions prevent its imaginary part, $\int_{0}^{t}d\tau\int_{0}^{\tau}d\sigma  V_{-}(\tau) \gamma_i(\tau-\sigma) V_{+}(\sigma) \neq \frac{1}{2} \int_{0}^{t}d\tau\int_{0}^{t}d\sigma V_{-}(\tau) \gamma_i(\tau-\sigma) V_{+}(\sigma) $, from being interpreted as classic Gaussian noise since  $\gamma_i(\tau-\sigma) $ is an odd function $\gamma_i(\tau-\sigma) = -\gamma_i(\sigma-\tau)$ and $V_{-}(\tau) V_{+}(\sigma) = V_{+}(\sigma)V_{-}(\tau)$ does not hold. As a result, constructing a real-valued Gaussian process with proper temporary and spatial structures to match the real-time Feynman-Vernon influence functional is not a trivial but a possible direction with possible extensions of Gaussian processes. Previously, complex-valued Gaussian processes were constructed\cite{chencaosilbey} by me and my co-workers to deconvolute the real-time Feynman-Vernon influence functional. However the Monte Carlo sampling of the complex-valued Gaussian process still has the sign problem and is not trivial numerically\cite{mak}. More novel and interesting construction approaches are needed and will be pursued in the future.

\section{Coupling Matrix, Local and Nonlocal Phonon}\label{coupling}
In the system-bath Hamiltonian, the phonon can be local or non-local in the interaction part.
The choice of nonlocal phonon will lead to
\begin{equation}
\hat{H}_I(q,X) = \bigg ( \sum_i \sum_j V_{ij} \vert i \rangle \langle j \vert \bigg ) \hat{\mathbf{X}} ,
\end{equation}
and the choice of local phonon
\begin{equation}
\hat{H}_I(q,X) = \sum_i \sum_j \bigg ( \vert i \rangle \langle j \vert \hat{\mathbf{X}}_{ij} \bigg) .
\end{equation}
For the simplification, it assumes in this paper that the set of modes in $\hat{\mathbf{X}}_{ij}$ shares no common element with $\hat{\mathbf{X}}_{kl}$ in which $i \neg k$ or $j \neq l$ or $i \neq k$ and $j \neq l$ for the local phonon. If the set of modes in $\hat{\mathbf{X}}_{ij}$ shares some common elements with $\hat{\mathbf{X}}_{kl}$, then there will be correlation between $V_{ij}(t)$ and $V_{kl}(t)$ over the path. If the set of modes $ \hat{\mathbf{X}}_{ij}$ is identical with $\hat{\mathbf{X}}_{kl}$, then we recover the the nonlocal phonon.

For the nonlocal phonon, one Gaussian process will be needed since there is only one influence function in the path integral formalism in Eq.~\ref{einfl}. For the local phonon, independent Gaussian processes are needed for each matrix element, $\vert i \rangle \langle j \vert$ since there are multi-state dependent influence functionals\cite{weiss,cnumber},
\begin{equation}\label{einfl2}
S_{infl}^{(E)}  =  \int_0^{\hbar \beta} \mu V_{ij}[q(\tau)] ^2 - \int_0^{\hbar \beta} d\tau \int_0^{\tau} d\sigma K_{ij}(\tau-\sigma) V_{ij}[q(\tau)] V_{ij}[q(\sigma)],
\end{equation}
given that the set of $\hat{\mathbf{X}}_{ij}$ shares no common element with $\hat{\mathbf{X}}_{kl}$.
The derivation can be found in the reference\cite{weiss} about the state dependent friction and the article about quantum non-Markovian process\cite{cnumber}.

\subsection{Coupling Matrix and Nonlocal Phonon}\label{nonlocal}

In this subsection, the diagonal coupling matrix $\hat{V}$, $\it{i.e.}$ $V_{ij}=0,\; i\neq j$, is considered. The famous case is the spin-boson model in which $V_{11}=-V_{22}$. It is well known that the diagonal coupling matrix will lead to the dephasing in reduced quantum systems.
Without loss of generality, I use two level systems to elaborate the ideas.
The detailed derivation for two level systems (spin-boson models) can be found in reference \cite{stefan}.
The $\hat{V}$ can be decomposed to be,
\begin{equation}\label{interference}
\hat{V} = \hat{V}_1+\hat{V}_2,
\end{equation}
where
\begin{equation}
\hat{V}_1=
\begin{pmatrix}
V_{11}   & 0 \\
0   & 0
\end{pmatrix},
\end{equation}
and
\begin{equation}
\hat{V}_2=
\begin{pmatrix}
0   &  0 \\ \nonumber
0 & V_{22}
\end{pmatrix}.
\end{equation}
Consequently,
with this  decomposition, the influence functional can expressed as,
\begin{eqnarray}\label{kernelmat2}
V(\tau)V(\sigma)K(\tau-\sigma)=V_1(\tau)V_1(\sigma)K(\tau-\sigma)+V_2(\tau)V_2(\sigma)K(\tau-\sigma) \\ \nonumber
+V_1(\tau)V_2(\sigma)K(\tau-\sigma)+V_2(\tau)V_1(\sigma)K(\tau-\sigma),
\end{eqnarray}
where $V_1(t)$ is the continuous limit\cite{segal,stefan,friesner,winter} of $\langle q(t) \vert \hat{V}_1 \vert q(t+dt) \rangle$ and $V_2(t)$ is the continuous limit of $\langle q(t) \vert \hat{V}_2 \vert q(t+dt) \rangle$.

In order to elaborate my idea below, I define
\begin{equation}
\hat{V}_1^e =
\begin{pmatrix}
1   &  0 \\ \nonumber
0 & 0
\end{pmatrix},
\end{equation}
and
\begin{equation}
\hat{V}_2^e =
\begin{pmatrix}
0   &  0 \\ \nonumber
0 & 1
\end{pmatrix}.
\end{equation}
Therefore,
\begin{eqnarray}
&&\exp\big \{ V_1(\tau)V_1(\sigma)K(\tau-\sigma)+V_2(\tau)V_2(\sigma) K(\tau-\sigma) \\ \nonumber
&&+V_1(\tau)V_2(\sigma)K(\tau-\sigma)+V_2(\tau)V_1(\sigma)K(\tau-\sigma) \big \} = \\ \nonumber
&&\exp\big \{ V_1^e(\tau)V_1^e(\sigma)V_{11}^2K(\tau-\sigma)+V_2^e(\tau)V_2^e(\sigma)V_{22}^2 K(\tau-\sigma) \\ \nonumber
&&+V_1^e(\tau)V_2^e(\sigma)V_{11}V_{22}K(\tau-\sigma)+V_2^e(\tau)V_1^e(\sigma)V_{22}V_{11}K(\tau-\sigma) \big \} = \\ \nonumber
&&\bigg \langle \exp \big \{ V_1^e(\tau) \xi_1^e(\tau) + V_2^e(\tau) \xi_2^e(\tau) \big \} \bigg \rangle_{\xi_1^e(\tau),\xi_2^e(\tau)},
\end{eqnarray}
in which the joint covariance matrix of $\xi_1$ and $\xi_2$ is
\begin{equation}\label{join}
\begin{pmatrix}
\langle \xi_{1}^e (\tau) \xi_{1}^e (\sigma) \rangle & \langle \xi_{1}^e (\tau) \xi_{2}^e (\sigma) \rangle  \\
\langle \xi_{2}^e (\tau) \xi_{1}^e (\sigma) \rangle & \langle \xi_{2}^e (\tau) \xi_{2}^e (\sigma) \rangle
\end{pmatrix},
\end{equation}
where $\langle \xi_{1}^e (\tau) \xi_{1}^e (\sigma)=V_{11}^2K(\tau-\sigma)$, $\langle \xi_{2}^e (\tau) \xi_{2}^e (\sigma) \rangle=V_{22}^2 K(\tau-\sigma)$ and
$\langle \xi_{1}^e (\tau) \xi_{2}^e (\sigma) \rangle=\langle \xi_{2}^e (\tau) \xi_{1}^e (\sigma) \rangle=V_{11}V_{22}K(\tau-\sigma)$.

This expression tell us that the origin of damping is from the discrepancy of $V_{11}$ and $V_{22}$, $\it{i.e.}$ the couplings between individual energy levels and phonon should be different from each other.
Clearly, Eq.~\ref{join} shows that if $V_{11}=V_{22}$, the Gaussian fluctuations, $\xi_1^e(t)$ and $\xi_2^e(t)$ will be perfectly correlated since $V_{11}^2=V_{22}^2=V_{11}V_{22}$. As a result, the system will be in perfect coherence. On the other hand, for the spin-boson model, $V_{11}=-V_{22}$, the Gaussian fluctuations on level 1 and 2 will be perfectly anti-correlated since $V_{11}^2=V_{22}^2=-V_{11}V_{22}$. The coherence will be in destruction. For the general case $V_{11}\neq V_{22}$, the coherence will be in destruction as well in different ways. In the results section, I will take examples to explain this ideas.
According to the construction procedure,  we can obtain the effective interaction, $\hat{V}\circ \boldsymbol\xi(t)$, in which $\circ$ is element-wise multiplication 
\begin{equation}
\hat{V}\circ \boldsymbol\xi(t)=
\begin{pmatrix}
V_{11}*\xi_{11}   &  0 \\
0  & V_{22}*\xi_{22}
\end{pmatrix},
\end{equation}
where 
\begin{equation}
\boldsymbol{\xi(t)}=
\begin{pmatrix}
\xi_{11}(t)   &  0 \\
0  & \xi_{22}(t)
\end{pmatrix}.
\end{equation}

\subsection{Coupling Matrix and Local Phonon}\label{local}
For the local phonon, similar to the nonlocal phonon, we can construct the Gaussian process with the corresponding covariance matrix.
Given the structure of influence functional in Eq.~\ref{einfl2}, the following equality can be constructed, \begin{eqnarray}
&&\exp\big \{ V_1(\tau)V_1(\sigma) K(\tau-\sigma)+V_2(\tau)V_2(\sigma) K(\tau-\sigma) \\ \nonumber
&&\exp\big \{ V_1^e(\tau)V_1^e(\sigma)V_{11}^2K(\tau-\sigma)+V_2^e(\tau)V_2^e(\sigma)V_{22}^2 K(\tau-\sigma) \\ \nonumber
&=&\bigg \langle \exp \big \{ V_1^e(\tau) \xi_1(\tau) + V_2^e(\tau) \xi_2(\tau) \big \} \bigg \rangle_{\xi_1^e(\tau),\xi_2^e(\tau)},
\end{eqnarray}
in which
\begin{equation}
\begin{pmatrix}
\langle \xi_{1}^e (\tau) \xi_{1}^e (\sigma) \rangle & \langle \xi_{1}^e (\tau) \xi_{2}^e (\sigma) \rangle  \\
\langle \xi_{2}^e (\tau) \xi_{1}^e (\sigma) \rangle & \langle \xi_{2}^e (\tau) \xi_{2}^e (\sigma) \rangle
\end{pmatrix},
\end{equation}
where $\langle \xi_{1}^e (\tau) \xi_{1}^e (\sigma)=V_{11}^2K(\tau-\sigma)$, $\langle \xi_{2}^e (\tau) \xi_{2}^e (\sigma) \rangle=V_{22}^2 K(\tau-\sigma)$ and
$\langle \xi_{1}^e (\tau) \xi_{2}^e (\sigma) \rangle=\langle \xi_{2}^e (\tau) \xi_{1}^e (\sigma) \rangle=0$.
According to the construction procedure, we can obtain the similar effective interaction as in the presvious Subsection~\ref{nonlocal}, $\hat{V}\circ \boldsymbol\xi(t)$, in which $\circ$ is element-wise multiplication and
\begin{equation}
\boldsymbol{\xi(t)}=
\begin{pmatrix}
\xi_{11}(t)   &  0 \\
0  & \xi_{22}(t)
\end{pmatrix}.
\end{equation}
If $\hat{V}$ is a full matrix,
\begin{equation}
\boldsymbol{\xi(t)}=
\begin{pmatrix}
\xi_{11}(t)   &  \xi_{12}(t) \\
\xi_{12}(t)  & \xi_{22}(t)
\end{pmatrix}.
\end{equation}

\section{REDM and Stochastic Matrix Multiplication Scheme}\label{citpi}
For the Euclidean imaginary time influence functional defined in Eqs.~\ref{Ham} and ~\ref{einflu}, if $V(q)$ is a diagonal matrix, two independent Gaussian noises are required \cite{chencaosilbey} and can be sampled easily with the Cholesky decomposition method\cite{cholesky,chencaosilbey}. In terms of the path integral formalism, REDM can be expressed as\cite{weiss},
\begin{equation}\label{redm}
\rho_{\beta} (q'',q')= \frac {1}{Z} \int_{q{0}=q'}^{q(\hbar \beta)=q''} D q(\cdot) \exp \bigg \{ -S_{eff}^{E} [q(\cdot)] /\hbar \bigg\},
\end{equation}
where $S_{eff}^{E}[q(\cdot)] = S_{S}^{E}[q(\cdot)] + S_{infl}^{E}[q(\cdot)] $, the effective Euclidean action.
For the nonlocal phonon, the Euclidean influence functional can be treated as the Gaussian moment generating function. Since  the Gaussian process, $\xi(t)$, is independent of the evolution of system dynamics, REDM can be re-written in terms of the average of stochastic matrix multiplication scheme,
\begin{equation}\label{redm1}
\rho_{\beta} (q'',q')= \frac {1}{Z} \bigg \langle \int_{q(0)=q'}^{q(\hbar \beta)=q''} D q(\cdot) \exp \bigg \{ \big \{ -S_{S}^{E} [q(\cdot)] - \int_0^{\hbar \beta} d \tau \mu V[q(\cdot)] ^2  - \int_0^{\hbar \beta} d \tau \xi(\tau) V(q(\cdot)) \big \}/\hbar \bigg\} \bigg \rangle_{\xi(t)},
\end{equation} or in terms of the compact effective Hamiltonian,
\begin{equation}\label{redm2}
\rho_{\beta} (q'',q')= \frac {1}{Z}
\bigg \langle  \exp( -\beta \hat{H}_{eff} ) \bigg \rangle_{\xi(t)},
\end{equation}
which is the continuous limit of the following stochastic matrix product
\begin{eqnarray}
&&\int d\hat{\boldsymbol{\xi}} f(\hat{\boldsymbol{\xi}}) \sum_{q_0}\sum_{s_1}\cdots\sum_{q_{N-3}}\sum_{q_{N-2}}\sum_{q_{N-1}}
\langle q' \vert \exp
\big\{ -  \hat{H}_{eff}   d\beta \big\} \vert q_{N-1} \rangle \\ \nonumber
&&
\langle q_{N-1} \vert \exp
\big\{ -  \hat{H}_{eff}   d\beta \big\} \vert q_{N-2} \rangle
\langle q_{N-2} \vert \exp
\big\{ -  \hat{H}_{eff}   d\beta \big\} \vert q_{N-3} \rangle \\ \nonumber
&&\cdots
\langle q_1 \vert \exp
\big\{ -  \hat{H}_{eff}   d\beta \big\} \vert q_0 \rangle
\langle q_0 \vert \exp
\big\{ -  \hat{H}_{eff}   d\beta \big\} \vert q'' \rangle
\end{eqnarray}
in which,
\begin{equation}
\hat{H}_{eff}=\hat{H}_S(q)+\mu \hat{V}(q)\circ \hat{V}(q)+\hat{V}(q) {\xi}(t),
\end{equation}
\begin{equation}
f(\hat{\boldsymbol{\xi}})=\frac{1}{\sqrt{(2\pi)^{N+1} \vert \Sigma \vert }} \exp(-\frac{1}{2} \hat{\boldsymbol{\xi}^{\dagger}} {\Sigma} \hat{\boldsymbol{\xi}}),
\end{equation}
$\hat{\boldsymbol{\xi}}=[\xi(0),\xi(t_0),\cdots,\xi(t_{N-2}),\xi(t_{N-1}),\xi(\beta \hbar)]$,
and $\sum_{q_0}\sum_{q_1}\cdots\sum_{q_{N-3}}\sum_{q_{N-2}}\sum_{q_{N-1}}$ is the summation over the discrete path on the time grid, $[\beta, t_{N-1},t_{N-2},t_{N-3}, \cdots t_1, t_0, 0]$.
The detailed discussion can be found in Appendix~\ref{drift}.
The covariance matrix, ${\Sigma}$ of the Gaussian process ${\xi}(t)$ can be defined as,
\begin{equation}\label{covm}
\boldsymbol{\Sigma}_i=\begin{pmatrix}
  \langle \xi(t_0)\xi(t_0)\rangle  & \langle\xi(t_0)\xi_i(t_1) \rangle& \cdots & \langle\xi(t_0)\xi(t_n) \rangle\\
  \langle \xi(t_1)\xi(t_0) \rangle& \langle\xi(t_1)\xi_i(t_1) \rangle& \cdots & \langle\xi(t_1)\xi(t_n) \rangle\\
  \vdots  & \vdots  & \ddots & \vdots  \\
  \langle\xi(t_n)\xi(t_0)\rangle & \langle\xi(t_n)\xi_i(t_1) \rangle& \cdots & \langle\xi(t_n)\xi(t_n)\rangle
 \end{pmatrix},
\end{equation}  where
$\langle\xi(t_i)\xi(t_j)\rangle = \langle\xi(t_j)\xi(t_i)\rangle = K(t_i-t_j)$ given the periodicity of $K(t)$ and its symmetry in its definition.
\section{Results}\label{results}
I consider the simplest dimer Hamiltonian of exciton-phonon coupling (spin-boson model) in which,
\begin{equation}
\hat{H}_S=\begin{pmatrix}
\epsilon_0 & J \\
J & \epsilon_1
\end{pmatrix},
\end{equation}
in which
\begin{equation}
\hat{H}_S=\epsilon \hat{\sigma}_z + J \hat{\sigma}_x,
\end{equation}
$\epsilon_0=-60$ cm $^{-1}$, $\epsilon_1=60$ cm $^{-1}$, $\epsilon=60$ cm $^{-1}$, and $J=60$ cm $^{-1}$,
and the diagonal coupling matrix $\hat{V}$ is defined as,
\begin{equation}
\hat{V}=\begin{pmatrix}
-1 & 0 \\
0 & 1
\end{pmatrix},
\end{equation}
which is equal to $\hat{\sigma}_z$ in the spin boson model.
An ohmic spectral density with an exponential cutoff,
\begin{equation}\label{ohmic}
J(\omega)=\eta \omega \exp(-\omega/\omega_c)
\end{equation}
is used in the dimer model to determine the coefficients of coupling, where $\omega_c$ is a cutoff frequency and the amplitude $\eta$ is a classically measurable friction coefficient of the bath which, in some context, reflects the amplitudes of the nuclear displacements.
In the following three subsections, I will discuss the kernel of imaginary time influence functional and the computational results of REDM.

\subsection{Kerenl of Ohmic Spectral Density and Monte Carlo Reproduction}
The kernel $K(\tau)$ in Eq.~\ref{kernel} and $\mu$ in Eq~\ref{einflu} are two key factors in deciding REDM. For the kernel $K(\tau)$, temperature $T$ in $\beta$, and $\eta$ and $\omega_{c}$ in the Ohmic spectral density play an important role in REDM. Eq.~\ref{covm} shows how the covariance matrix of Gaussian process is determined by the kernel $K(\tau)$ with the discretization on the time grid.

\begin{figure}
\includegraphics[width=6in]{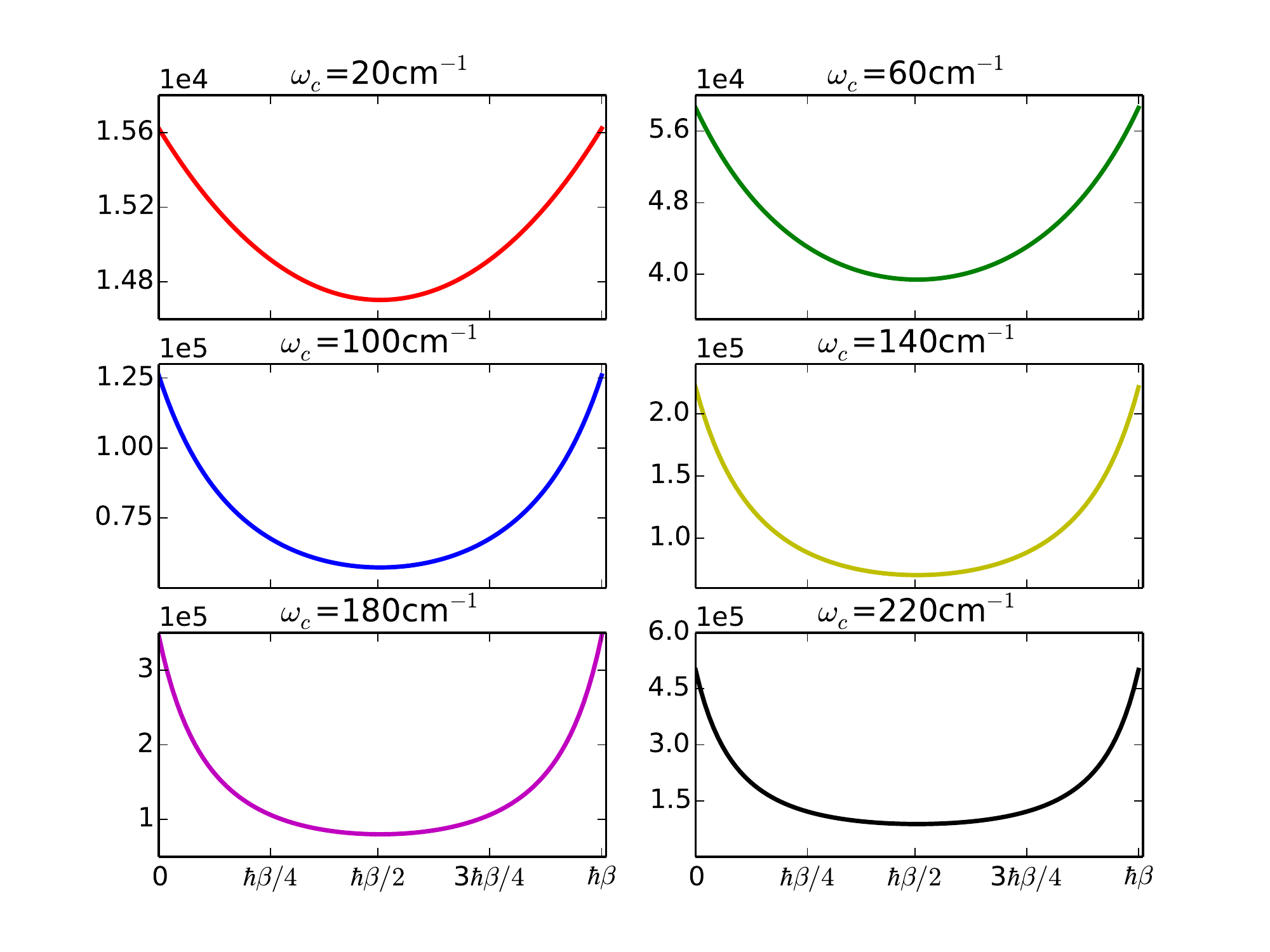}
\caption{The window of the kernel from $0$ to $ \hbar \beta$ is presented for $\omega_c=$ 20, 60, 100, 140, 180, 220 $cm^{-1}$  (T=300k and $\eta=J/2$).}\label{kernelomegac}
\end{figure}

Figures~\ref{kernelomegac} shows the window of $K(\tau)$ from $0$ to $\hbar \beta$ for different $\omega_c$. It shows that the larger $\omega_c$, the wider the kernel becomes.
Since kernel's dependence on temperatures and $\eta$ is similar to $\omega_c$, it will not be presented without the loss of important information.

The Cholesky decomposition \cite{chencaosilbey} can be used to decompose the covariance matrix of a temporarily and spatially correlated Gaussian process to generate discreet sample Gaussian vectors, $\boldsymbol{\xi}(t_i)=[\xi^1(t_i),\xi^2(t_i),\xi^3(t_i),\cdots \xi^M(t_i)]$ of the Gaussian process by sampling independent normal distribution random variables with the Monte Carlo method
on the time grid, ${t_0,t_1,t_2,\cdots,t_N}$ in which $t_0=0$ and $t_N=\beta \hbar$.  In the Table~\ref{tab1}, I show the comparison of the analytic value of kernel, $K(\tau-\sigma)$, and the sample covariance,
\begin{equation}
\langle \xi(t_i) \xi (t_j)  \rangle=\frac{1}{M-1} \sum_{k=1}^M [\xi^k(t_i)-\bar{\xi}(t_i)][\xi^k(t_j)-\bar{\xi}(t_j)],
\end{equation}
where
\begin{equation}
\bar{\xi}(t_i)= \frac{1}{M} \sum_{k=1}^M \xi^k(t_i).
\end{equation}
The parameters are $\eta=J/2$, $\omega_c=100$, $T=300k$ and $N=10$. In the table, it is clearly shown that the Monte Carlo method with the Cholesky decomposition method can reproduce the symmetry of the kernel which is important for the calculation of REDM since the two boundary points $0$ and $\beta \hbar$ are the constraints for random processes.  The most famous example is the Brownian bridge for the Brownian motion\cite{brownian}. Table~\ref{tab1} shows that the sample covariances converge to the analytic ones when the number of trajectories increases.
\begin{table}
\begin{tabular}{l*{6}{c}r}
Corr Fun      \;  & \; $\langle \xi(0) \xi(0) \rangle$ \; & \; $\langle \xi(0) \xi(\frac{\beta \hbar}{4}) \rangle$ \; & \; $\langle \xi(0) \xi(\frac{\beta \hbar}{2}) \rangle$
 \; & \; $\langle \xi(0) \xi(\frac{3 \beta \hbar }{4}) \rangle$ \; & \; $\langle \xi(0) \xi(\beta \hbar ) \rangle$ \\
\hline
Analytic         \; & \;125914.37 \; & \; 67871.98 \; & \; 57332.21\;  &\; 67871.98 \;& \; 125914.37  \\
1M Trajs         \; & \;125813.95 \; & \; 67725.39 \; & \; 57146.39\;  &\; 67881.79 \;& \; 125813.95   \\
10M Trajs         \; & \;125902.33 \; & \; 67824.01 \; & \; 57305.48\;  &\; 67810.65 \;& \; 125902.33   \\
\end{tabular}
\caption{Comparision of analytic kernel and numerical sample covariance}
\label{tab1}
\end{table}

In the following subsections, I calculate the normalized REDM for different temperatures and parameters of Ohmic spectral densities to compare with the corresponding Boltzmann distribution,

\subsection{Temperature Dependence}\label{temp-dep}
Given the Ohmic spectral density of Eq.~\ref{ohmic}, the friction term, $\mu$ in Eq.~\ref{einfl} has the expression,
$\mu=2 \eta \omega_c / \pi$, which is independent of temperatures. In this calculations, we  choose $\eta=J/2$ and $\omega_c=100 cm^{-1}$ for the Ohmic spectral density.
For the two level system, the matrix elements of the normalized REDM, $\frac{Tr_{\beta}(e^{-\beta H})}{Tr(e^{-\beta H})}$, are: $\text{normalized}\;  \rho_{11} = \frac{\rho_{11}}{\rho_{11}+\rho_{22}}$, $\text{normalized}\;  \rho_{22} = \frac{\rho_{22}}{\rho_{11}+\rho_{22}}$ and $\text{normalized}\;  \rho_{12} = \frac{\rho_{12}}{\rho_{11}+\rho_{22}}$.

\begin{figure}
\includegraphics[width=6in]{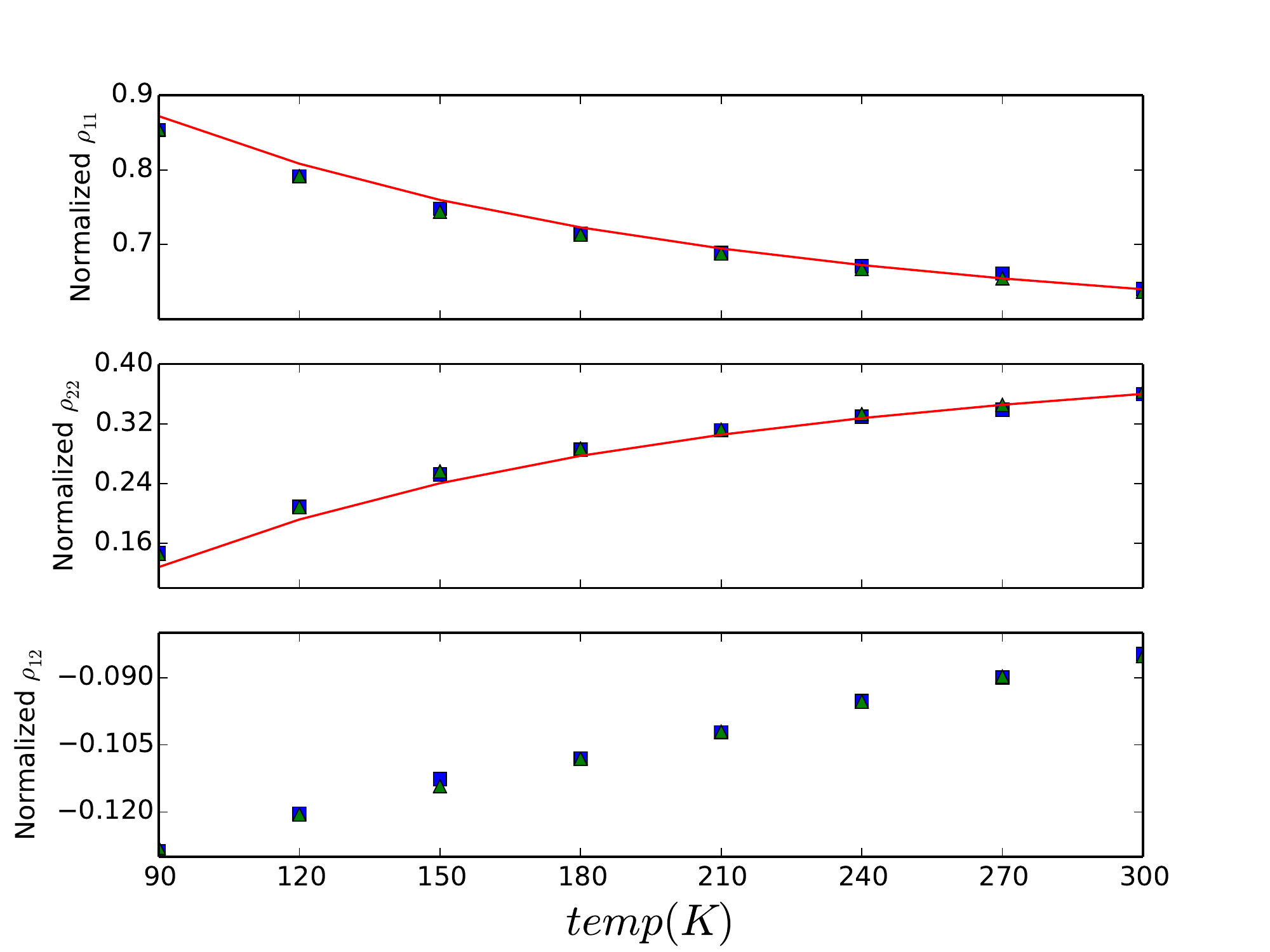}
\caption{Panel (a) shows that normalized $\rho_{11}$ changes with temperatures; Panel (b) shows that normalized $\rho_{22}$ with temperature; and Panel (c) shows that normalized $\rho_{12}$  with temperatures. The solid red line in Panels (a) and (b) is the Boltzmann equilibrium weights with respect to different temperatures (For blue squares, 40000 trajectories are used; for green triangles, 80000 trajectories. 300 time grid points are used )}\label{tempn}
\end{figure}

\begin{figure}
\includegraphics[width=6in]{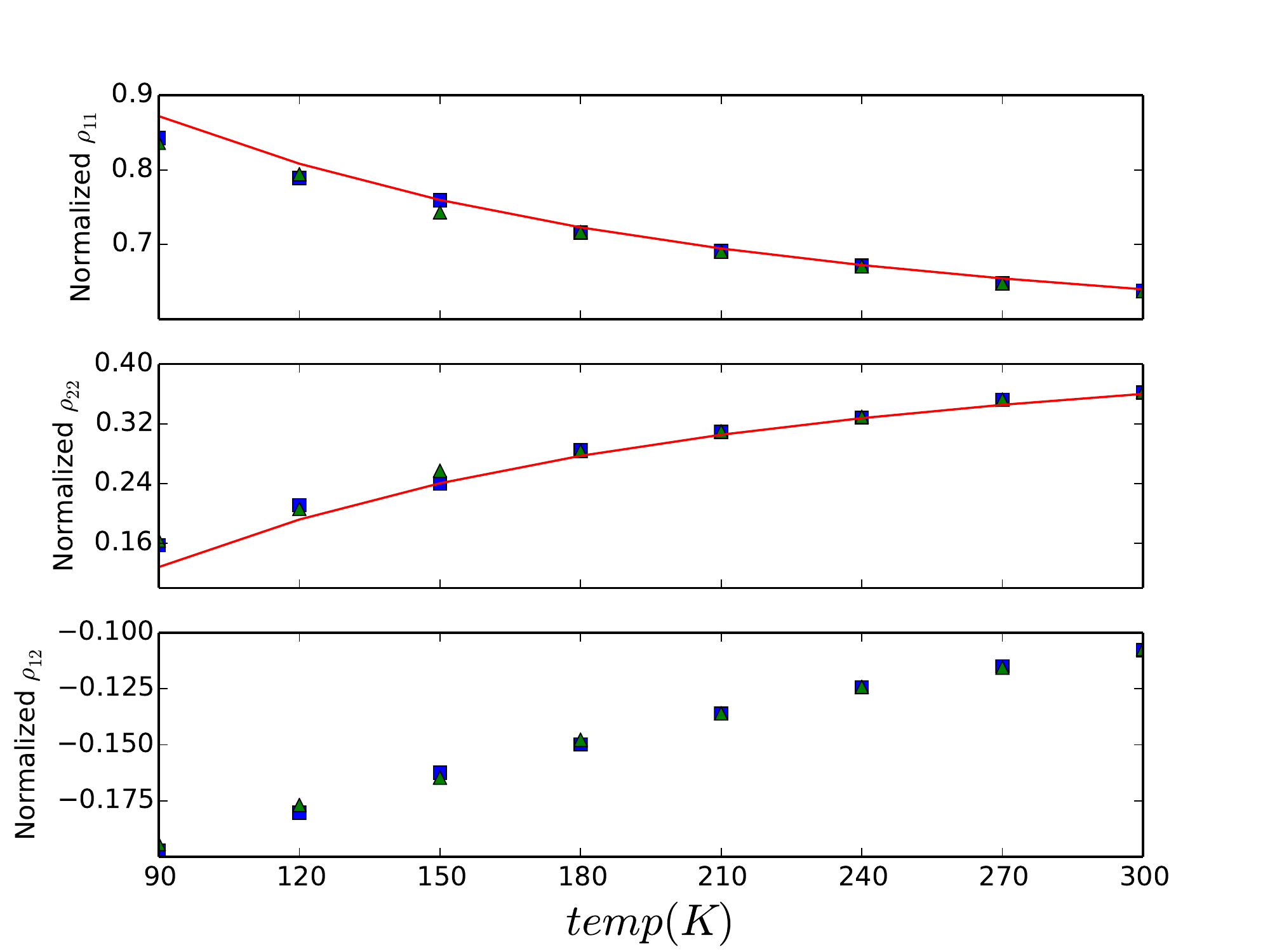}
\caption{Panel (a) shows that normalized $\rho_{11}$ changes with temperatures; Panel (b) shows that normalized $\rho_{22}$ with temperatures; and Panel (c) shows that normalized $\rho_{12}$  with temperature. The solid red line in Panels (a) and (b) is the Boltzmann equilibrium weights with respect to different temperatures (For blue squares, 40000 trajectories are used; for green triangles, 80000 trajectories. 300 time grid points are used )}\label{tempn-ind}
\end{figure}

In this subsection, I want to compare the results of two different couplings I discussed previously in subsections~\ref{nonlocal} and ~\ref{local}: 1. the coupling to nonlocal phonon, and 2. local phonon.
Figure ~\ref{tempn} present the matrix elements of the normalized REDM at different temperatures for the coupling to  nonlocal phonon. In Figure~\ref{tempn}, I show the comparison between the normalized $\rho_{11}$ and $\rho_{22}$ and the Boltzmann equilibrium distribution (the solid red lines). The figure clearly shows that the normalized $\rho_{11}$ and $\rho_{22}$ generally agree with the Boltzmann distribution for the high temperatures. However, for the low temperatures, the figure clearly shows that the normalized $\rho_{11}$ and $\rho_{22}$ are slightly off from  the Boltzmann distribution due to the quantum tunnelling. Interestingly,
the normalized coherence $\rho_{12}$ is not zero and decreases with temperatures in terms of its absolute values.
Figure~\ref{tempn-ind} present the matrix elements of normalized REDM at different temperatures for the coupling to local phonon. The normalized population $\rho_{11}$ and $\rho_{22}$ calculated from the local and nonlocal phonons are consistent with each other for the high temperatures. However, they have different normalized coherences. The results seem to imply that local and nonlocal phonons have different impacts on the quantum tunnelling at the low temperature. The local phonon coupling will lead to larger (in terms of absolute values) normalized coherence than the nonlocal phonon one because the anti-correlation implied by the spin-boson model discussed in subsection~\ref{nonlocal}.

On the side note, in both Figures~\ref{tempn} and ~\ref{tempn-ind}, the red lines corresponds to the Boltzmann equilibrium distribution of energy level 1 which is $\frac{e^{-\beta \epsilon_1 }}{e^{-\beta \epsilon_1 }+ e^{-\beta \epsilon_2 }}$ and energy level 2, $\frac{e^{-\beta \epsilon_2 }}{e^{-\beta \epsilon_1 }+ e^{-\beta \epsilon_2 }}$ for the different temperatures used in the plots.


\subsection{Spectral Density Parameter Dependence}\label{sdpd}

\begin{figure}
\includegraphics[width=6in]{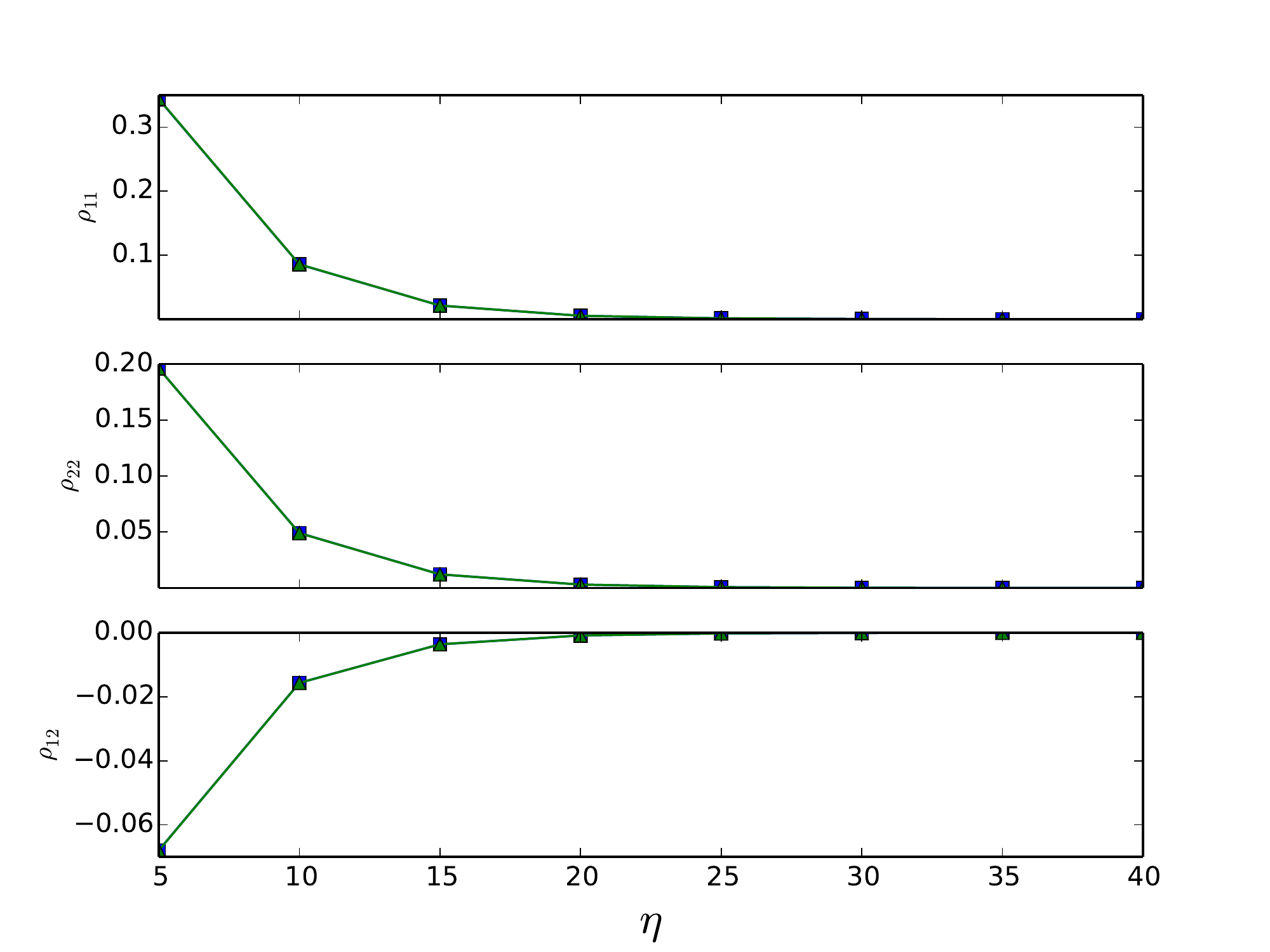}
\caption{Panel (a) shows the change of $\rho_{11}$ with the respect to $\eta$; Panel (b) shows the change of $\rho_{22}$ with respect to $\eta$; and Panel (c) shows the change of the $\rho_{12}$ with respect to $\eta$. The remaining parameters $\omega_c=100 cm^{-1}$ and $T=300K$ (For squares, 40000 trajectories are used; for triangles, 80000 trajectories are used. 300 time grid points are used )}\label{eta}
\end{figure}

\begin{figure}
\includegraphics[width=6in]{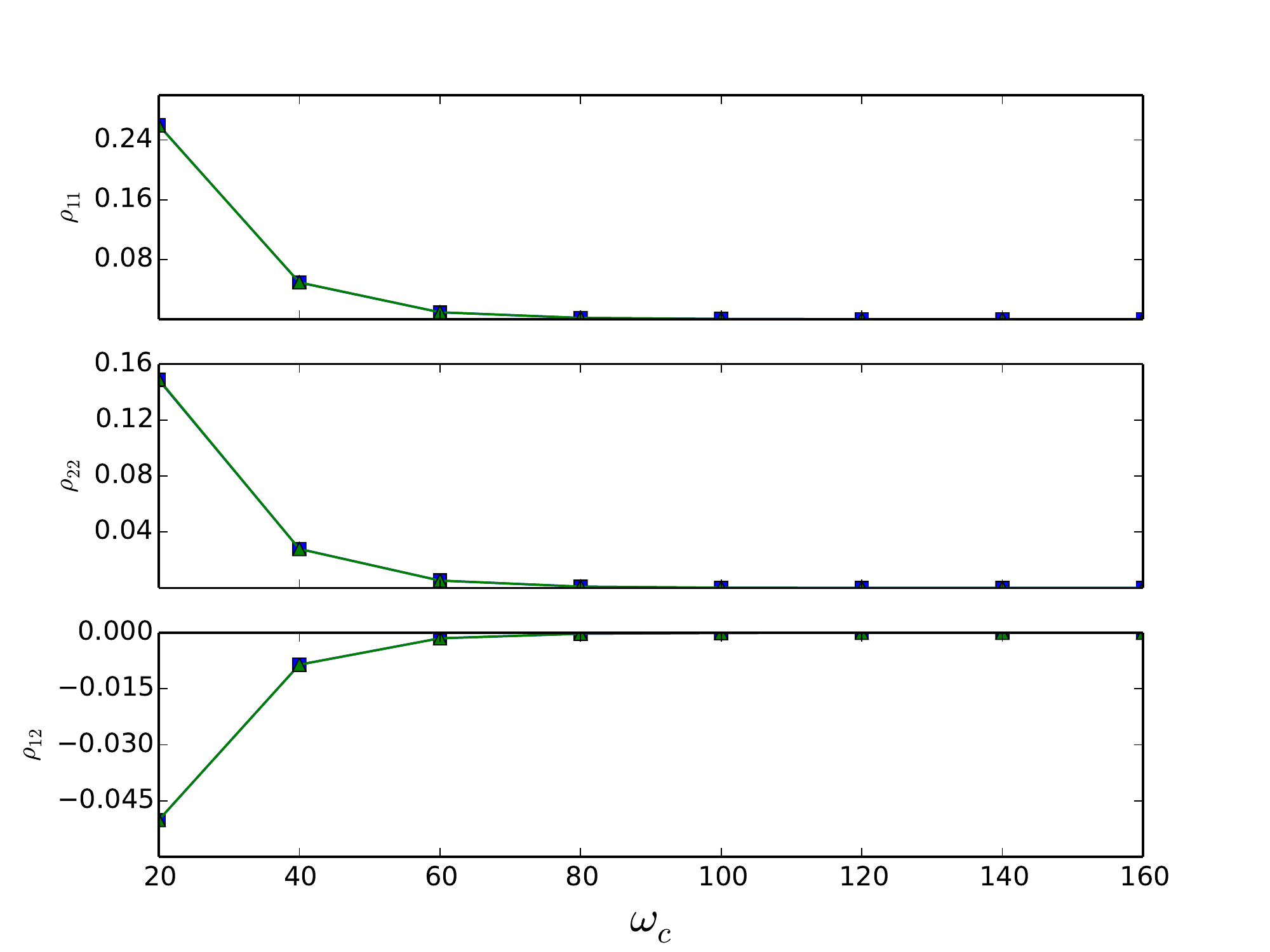}
\caption{Panel (a) shows the change of $\rho_{11}$ with respect to $\omega_c$; Panel (b) shows the change of $\rho_{22}$ with respect to $\omega_c$; and Panel (c) shows the change of $\rho_{12}$  with respect to $\omega_c$. The remaining parameters are $\eta=J/2$ and $T=300K$. (For squares, 40000 trajectories are used; for triangles, 80000 trajectories are used. 300 time grid points are used )}\label{omegac}
\end{figure}

Figures ~\ref{eta} and ~\ref{omegac} clearly show that with the increase of $\eta$ and $\omega_c$, the REDM elements, $\rho_{11}$, $\rho_{22}$, and $\rho_{12}$ will decrease since $\mu$ increases with $\eta$ and $\omega_c$ and the surfaces under the curves of kernels. This picture shows that in total, when $\eta$ and $\omega_c$ get bigger, more energy will reside in thermal reservoir. In Figure~\ref{eta}, the parameters are $\omega_c=100$cm$^{-1}$ and $T=300K$; and In Figure~\ref{omegac}, the parameters $\eta=J/2$ and $T=300K$

\begin{figure}
\includegraphics[width=6in]{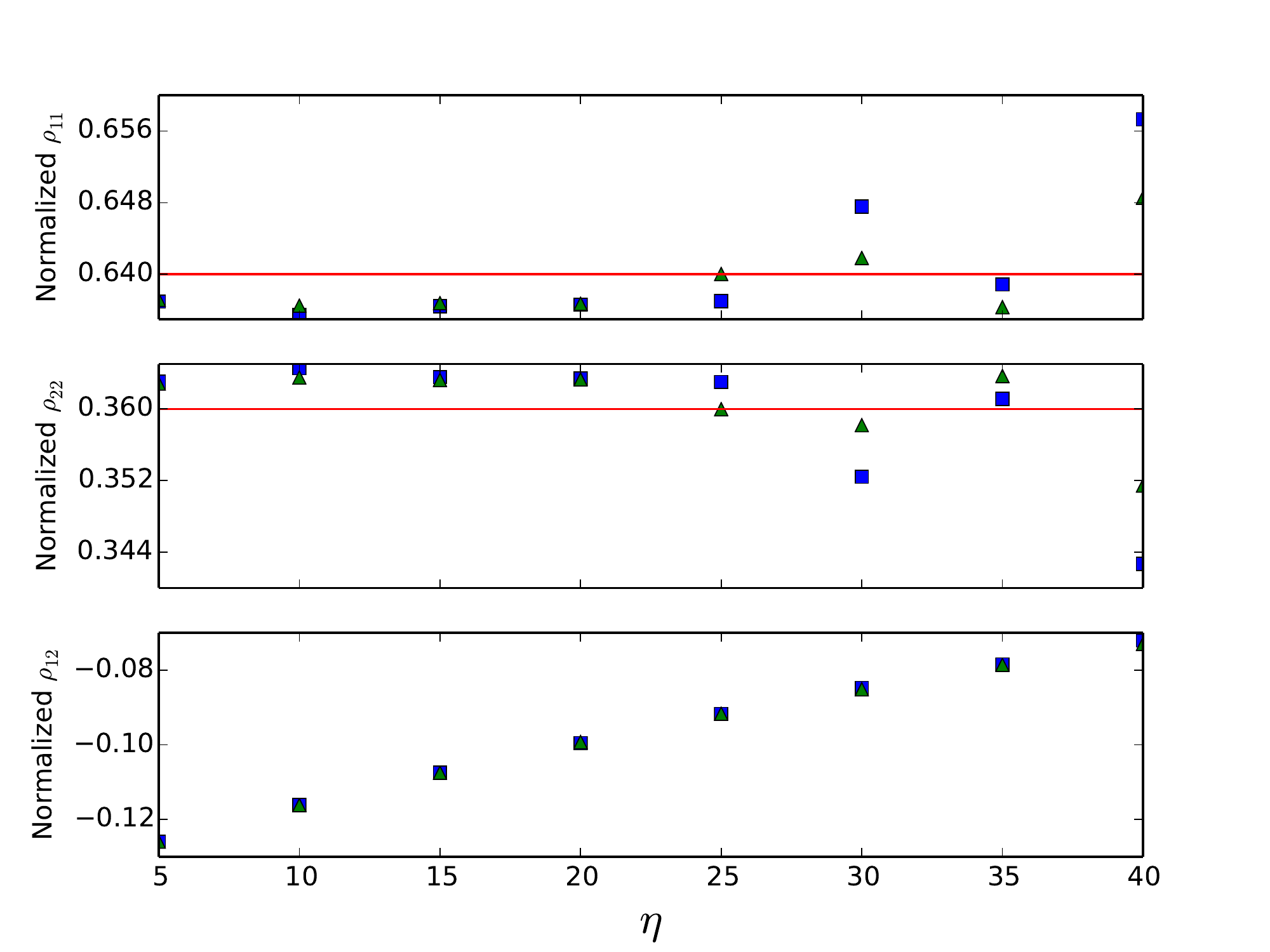}
\caption{Panel (a) shows the change of the normalized $\rho_{11}$  with respect to $\eta$; Panel (b) shows the change of the normalized $\rho_{22}$ with respect to $\eta$; and Panel (c) shows the change of the normalized $\rho_{12}$  with respect to $\eta$. The solid red line in Panels (a) and (b) is the Boltzmann distribution at 300K. The remaining parameters $\omega_c=100 cm^{-1}$ and $T=300K$ (For squares, 40000 trajectories are used; for triangles, 80000 trajectories are used. 300 time grid points are used )}\label{etan}
\end{figure}

\begin{figure}
\includegraphics[width=6in]{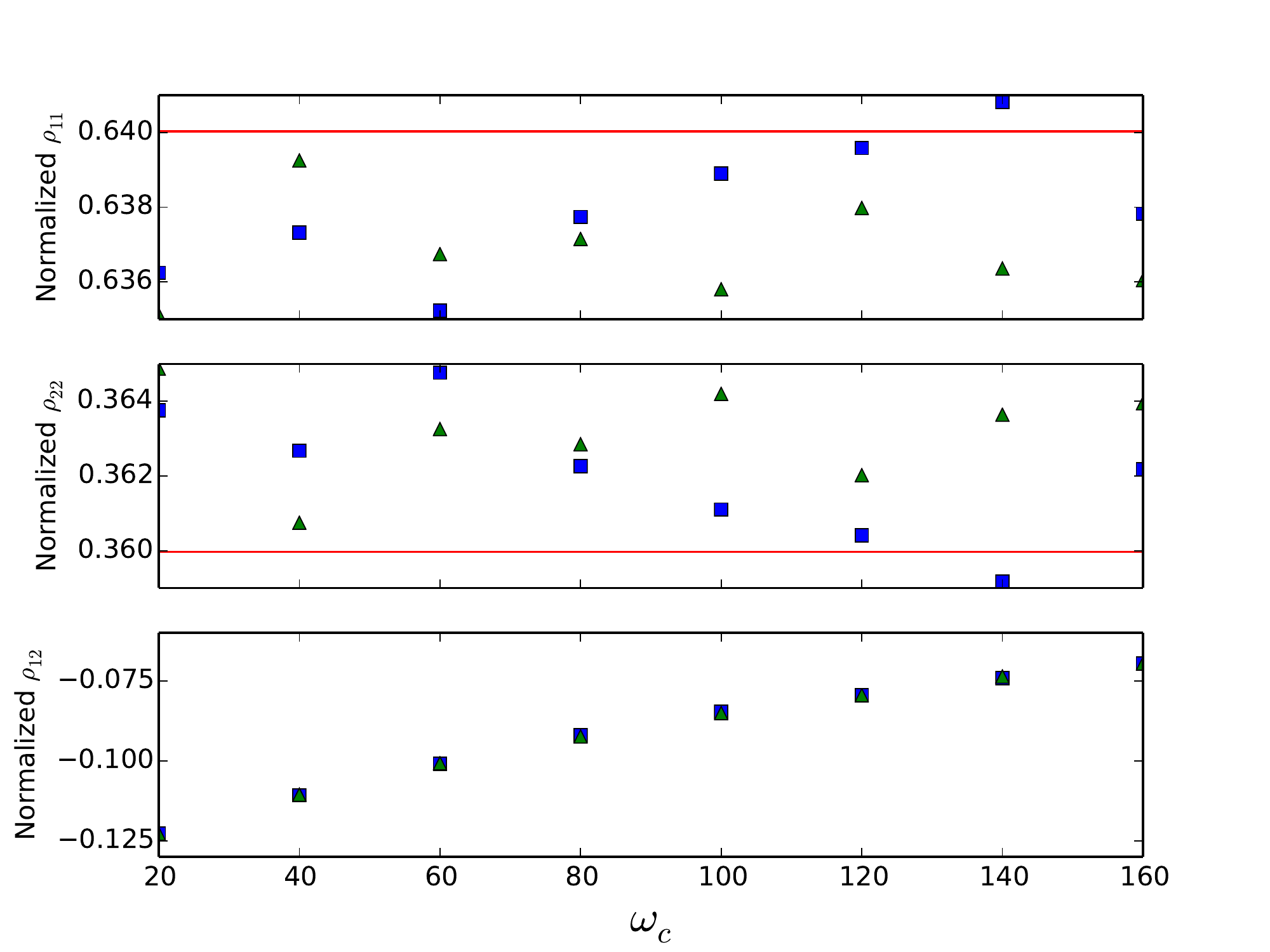}
\caption{Panel (a) shows the change of the normalized $\rho_{11}$  with respect to $\omega_c$; Panel (b) shows that normalized $\rho_{22}$ with respect to $\omega_c$; and Panel (c) shows that normalized $\rho_{12}$  with respect to $\omega_c$. The solid red line in Panels (a) and (b) is the Boltzmann distribution at 300K. The remaining parameters are $\eta=J/2$  and $T=300K$. (For squares, 40000 trajectories are used; for triangles, 80000 trajectories are used. 300 time grid points are used )}\label{omegacn}
\end{figure}

Slightly different from Figures~\ref{eta} and ~\ref{omegac}, Figure~\ref{etan} shows the plots of the normalized matrix elements of REDM with respect to $\eta$; Figure~\ref{omegacn} the normalized matrix elements of REDM with respect to $\omega_c$. In both figures, the red lines corresponding to the Boltzmann weight (distribution) of energy level 1, $\frac{e^{-\beta \epsilon_1 }}{e^{-\beta \epsilon_1 }+ e^{-\beta \epsilon_2 }}$ and energy level 2 $\frac{e^{-\beta \epsilon_2 }}{e^{-\beta \epsilon_1 }+ e^{-\beta \epsilon_2 }}$ are plotted to compare with the normalized populations in the normalized REDM. Clearly, for the coherence, $\rho_{12}$, the larger $\eta$, friction coefficient, the smaller equilibrium coherence; the larger $\omega$ (the width of spectral density), the smaller equilibrium coherence.

For the path integral Monte Carlo method, the error is roughly $\frac{1}{\sqrt{N}}$, where $N$ is the number of sampling points.
For Figures ~\ref{etan} and ~\ref{omegacn}, if we ignoring the numerical error due to the path integral propagation and step size, the errors for the calculations of 40000 sampling points (legend square) and 80000 (legend triangle) are around $\frac{1}{\sqrt{40000}}=0.005$ and $\frac{1}{\sqrt{80000}}=0.0035$. I also want to emphasize that the errors of normalized matrix elements of REDM are larger than the ones of matrix elements of REDM due to the ration of matrix elements and the trace. As a result, in Figures~\ref{etan} and ~\ref{omegacn}, the results are converged.

I discuss the effects of diagonal couplings with local and nonlocal phonons in the normalized REDM matrix elements using the simplest spin-boson model in subsections~\ref{temp-dep} and ~\ref{sdpd}.
The interesting effect of the off-diagonal coupling with phonon, $\it{i.e.}$ the fluctuations in electronic coupling, will be discussed in future papers.
Two interesting questions arising from the above calculations are: if we can solve the equation of motion of reduced density matrix (RDM) rigorously, will the steady states of open quantum systems have zero coherence? The current calculations show the existence of equilibrium coherences; what is the statistics of the steady state under quantum  path interferences with off-diagonal quantum fluctuations in the Hamiltonians, $\it{i.e.}$ will the non-Boltzmann distribution of the steady state show up in the existence of quantum path interferences/coherence.

\section{Concluding Remarks}\label{conrem}
In this paper, I discuss the Gaussian stochastic interpretations of imaginary and real time influence functionals from the constructive perspective, particularly the path ordering and non-commutativity of coupling matrix in the real-time Feynman-Vernon influence functional. I review the discrete Gaussian process moment generating function and characteristic function in the matrix representation.  With the comparison to the discrete Gaussian process moment generating function and characteristic function, I establish the conditions under which the influence functional can be reproduced as the average of real/imaginary-time stochastic evolution operator over the Gaussian trajectories. On top of the construction, I propose the stochastic matrix multiplication scheme to calculate REDM . For the simple spin-boson model, I discuss the coupling with the local and nonlocal phonons, the corresponding Gaussian processes and their structure of correlation and covariance matrices.  At the end, I simulate and calculate REDM of the spin-boson models for different temperatures, and parameters in Ohmic spectral densities with exponential cutoffs.

Many previous work based on perturbation methods shows that when $\hat{{V}}(q)$ is not diagonal matrix\cite{scully,coalson}, the non-Boltzmann long-time steady-state can be obtained. The future work will be two-fold: 1. to construct the stochastic real-time matrix product method to accommodate the path ordering of the Feynman-Vernon influence functional. 2. to study quantum path interferences/coherences when $V(q)$ is a general matrix for multi-level quantum systems and coupled to the local or nonlocal phonon. The extension of the current stochastic matrix multiplication scheme will enable us to gain rigorous insight to quantum path interferences in the real time dynamics of quantum subsystems and the break-down of Boltzmann equilibrium.

\appendix
\section{ Spin Boson Model and Influence Functional}\label{spinboson}

The standard spin boson model is $\hat{H}=\hat{H}_S + \hat{H}_I + \hat{H}_B$ where
\begin{equation}
\hat{H}_S =\frac{1}{2}(\epsilon \sigma_z + \Delta \sigma_x),
\end{equation}
and
\begin{equation}
\hat{H}_I=\hat{\sigma}_z  \hat{\mathbf{X}}.
\end{equation}

The derivation of influence function using the path integral will help understand Eq.~\ref{interference}. For the real time influence functional,
\begin{eqnarray}
\rho(q'',q',t)=\int d\mathbf{q}^+ \int d\mathbf{q}^- Tr_{B} \bigg\{   \langle q'' \vert \exp (-i \hat{H} dt) \vert q^+_{N-1} \rangle  \langle q^+_{N-1} \vert \exp (-i \hat{H} dt) \vert q^+_{N-2} \rangle
\cdots \\ \nonumber  \langle q^+_0 \vert \rho(0) \vert q^-_0 \rangle
\cdots  \langle q^-_{N-2} \vert \exp (-i \hat{H} dt) \vert q^-_{N-1} \rangle \langle q^-_{N-1} \vert \exp (-i \hat{H} dt) \vert q' \rangle  \bigg\},
\end{eqnarray}
where $dt=t/(N-1)$,  $\mathbf{q}^+$ is the forward trajectory, $\{q'',q^+_{N-1}, q^+_{N-2},\cdots, q^+_0\} $, $d\mathbf{q}^+ = d q^+_{N-1} dq^+_{N-2}\cdots dq^+_0 $, $\mathbf{q}^-$ is the backward trajectory, $\{q',q^-_{N-1}, q^-_{N-2},\cdots, q^-_0\} $ and $d\mathbf{q}^- = d q^-_{N-1} dq^-_{N-2}\cdots dq^-_0 $.
For the imaginary time influence functional,
\begin{eqnarray}
\rho^E(q'',q')=\int d\mathbf{q}  Tr_{B} \bigg\{   \langle q'' \vert \exp (- \hat{H} d\beta) \vert q_{N-1} \rangle \langle q_{N-1} \vert \exp (- \hat{H} d\beta) \vert q_{N-2} \rangle   \\ \nonumber
\cdots \langle q_2 \vert \exp (- \hat{H} d\beta) \vert q_0 \rangle \langle q_0 \vert \exp (- \hat{H} d\beta) \vert q' \rangle \bigg \}
\end{eqnarray}
where $d\beta = \beta /(N-1)$, $\mathbf{s}$ is the trajectory in the Euclidean path integral, $\{q'',q_{N-1}, q_{N-2},\cdots, q_0, q'\} $ and $d\mathbf{q} = d q_{N-1} dq_{N-2}\cdots dq_0 $. For the discrete quantum system like spin,  $\int d\mathbf{q}=\sum_{q_{N-1}}\sum_{q_{N-2}}\cdots\sum_{q_1}\sum_{q_0}$.
Given the structure of Hamiltonian,
\begin{eqnarray}
\rho(q'',q',t)=\int d\mathbf{q}^+ \int d\mathbf{q}^- Tr_{B} \bigg\{   \langle q'' \vert \exp (-i \hat{H} dt) \vert q^+_{N-1} \rangle  \langle q^+_{N-1} \vert \exp (-i \hat{H} dt) \vert q^+_{N-2} \rangle
\cdots \\ \nonumber  \langle q^+_0 \vert \rho(0) \vert q^-_0 \rangle
\cdots  \langle q^-_{N-2} \vert \exp (-i \hat{H} dt) \vert q^-_{N-1} \rangle \langle q^-_{N-1} \vert \exp (-i \hat{H} dt) \vert q' \rangle  \bigg\} = \\ \nonumber
 \int d\mathbf{q}^+ \int d\mathbf{q}^-  \langle q'' \vert \exp (-i \hat{H}_0 dt) \vert q^+_{N-1} \rangle  \langle q^+_{N-1} \vert \exp (-i \hat{H}_0 dt) \vert q^+_{N-2} \rangle
\cdots \\ \nonumber  \langle q^+_0 \vert \rho(0) \vert q^-_0 \rangle
\cdots  \langle q^-_{N-2} \vert \exp (-i \hat{H}_0 dt) \vert q^-_{N-1} \rangle \langle q^-_{N-1} \vert \exp (-i \hat{H}_0 dt) \vert q' \rangle I(\mathbf{q}^+ , \mathbf{q}^-,dt),
\end{eqnarray}
where
\begin{eqnarray}
I(\mathbf{q}^+ , \mathbf{q}^-)=\int d\mathbf{q}^+ \int d\mathbf{q}^- Tr_{B} \bigg\{   \langle q'' \vert \exp (-i \hat{H}_I dt) \vert q^+_{N-1} \rangle  \langle q^+_{N-1} \vert \exp (-i \hat{H}_I dt) \vert q^+_{N-2} \rangle
\cdots \\ \nonumber  \langle q^+_0 \vert \rho(0) \vert q^-_0 \rangle
\cdots  \langle q^-_{N-2} \vert \exp (-i \hat{H}_I dt) \vert q^-_{N-1} \rangle \langle q^-_{N-1} \vert \exp (-i \hat{H}_I dt) \vert q' \rangle  \bigg\}
\end{eqnarray}
and
\begin{eqnarray}
\rho^E(q'',q')=\int d\mathbf{q}  \langle s'' \vert \exp (- \hat{H}_0 d\beta) \vert q_{N-1} \rangle \langle q_{N-1} \vert \exp (- \hat{H}_0 d\beta) \vert q_{N-2} \rangle   \\ \nonumber
\cdots \langle s_2 \vert \exp (- \hat{H}_0 d\beta) \vert q_0 \rangle \langle q_0 \vert \exp (- \hat{H}_0 d\beta) \vert q' \rangle I(\mathbf{q},d\beta),
\end{eqnarray}
where
\begin{eqnarray}
I(\mathbf{q})=\int d\mathbf{q}Tr_{B} \bigg\{   \langle q'' \vert \exp (-i \hat{H}_I d\beta) \vert q_{N-1} \rangle  \langle q_{N-1} \vert \exp (-i \hat{H}_I d\beta) \vert q_{N-2} \rangle \\ \nonumber
 \cdots \langle q_1 \vert \exp (-i \hat{H}_I d\beta) \vert q_0 \rangle \langle q_0 \vert \exp (-i \hat{H}_I d\beta) \vert q' \rangle  \bigg\}
\end{eqnarray}
Following the standard practice in literature, when $dt \rightarrow 0$, for the standard spin-boson model \cite{winter},
\begin{eqnarray}
I(\mathbf{q}^+ , \mathbf{q}^-,dt) \rightarrow \exp \bigg \{ - \int_0^t d \tau\int_0^\tau d \sigma [f^+(\tau) -f^-(\sigma)] \gamma_r(\tau-\sigma) [f^+(\tau) -f^-(\sigma)] \\ \nonumber
+i \int_0^t d \tau\int_0^\tau d \sigma [f^+(\tau) -f^-(\sigma)] \gamma_i(\tau-\sigma) [f^+(\tau) + f^-(\sigma)]  \bigg \},
\end{eqnarray}
where $f^{\pm}(t)$ is a piecewise constant function\cite{segal} taking $1$ or $-1$, $\langle q^{\pm}(t) \vert \hat{\sigma}_z  \vert q^{\pm}(t+dt)\rangle$ in which $dt \rightarrow 0$.
Therefore,
\begin{eqnarray}
I(\mathbf{q}) \rightarrow \exp \bigg \{ -\int_0^{\hbar \beta} d\tau \int_0^{\tau} d\sigma K(\tau-\sigma) f(\tau) f(\sigma) \bigg \},
\end{eqnarray}
where $f^{\pm}(\beta)$ is a piecewise constant function, $\langle q^{\pm}(\beta) \vert \hat{\sigma}_z  \vert q^{\pm}(\beta+\epsilon)\rangle$ in which $\epsilon \rightarrow 0$.
In these expressions, I dropped the drift term in the influence functionals, $+ i \frac{\mu}{2} \int_0^t d\tau \big [ f^+(\tau)^2 -f^-(\tau)^2 \big ] $ and
$\int_0^{\hbar \beta} \mu f(\tau) ^2$.

For the extended spin-Boson model,
\begin{eqnarray}
I(\mathbf{q}^+ , \mathbf{q}^-,dt) \rightarrow \exp \bigg \{ - \int_0^t d \tau\int_0^\tau d \sigma [f_e^+(\tau) -f_e^-(\sigma)] \gamma_r(\tau-\sigma) [f_e^+(\tau) -f_e^-(\sigma)] \\ \nonumber
+i \int_0^t d \tau\int_0^\tau d \sigma [f_e^+(\tau) -f_e^-(\sigma)] \gamma_i(\tau-\sigma) [f_e^+(\tau) + f_e^-(\sigma)]  \bigg \},
\end{eqnarray}
where $f_e^{\pm}(t)=f^{\pm}(t)+f_o^{\pm}(t)$ and $f_o^{\pm}(t)$ is also a piecewise constant function taking $1$ or $-1$, $\langle q^{\pm}(t) \vert \hat{\sigma}_x  \vert q^{\pm}(t+dt)\rangle$. The same extension can be used for the imaginary time influence functional,
\begin{eqnarray}
I(\mathbf{q}) \rightarrow \exp \bigg \{ - \int_0^{\hbar \beta} d\tau \int_0^{\tau} d\sigma K(\tau-\sigma) f_e(\tau) f_e(\sigma) \bigg \},
\end{eqnarray}

\section{Rediscretization of Influence Functional}\label{drift}
The following derivation considering two-level systems with diabatic energy basis set $\vert 1 \rangle$ and $\vert 2 \rangle$.
The trick is to re-discretize  the influence functional in Eq.~\ref{einfl} over the path $\mathbf{q}$ with $V(t)$ as the continuous limit of $\langle q(t) \vert \hat{V} \vert q(t+d\beta )\rangle $. The details of proof can be found in the reference\cite{chencaosilbey}.
\begin{eqnarray}
I &\approx& \int d\hat{\boldsymbol{\xi}} f(\hat{\boldsymbol{\xi}}) \sum_{q_0}\sum_{q_1}\cdots\sum_{q_{N-2}}\sum_{q_{N-1}}  \\ \nonumber
&&
\exp
\bigg\{
- \big [  \mu\langle q'' \vert V \vert q_{N-1} \rangle ^2 + \langle q'' \vert \hat{V}  \vert q_{N-1} \rangle \xi(t_{N-1}) \big ] d\beta -
 \\ \nonumber &  &
\big [\mu \langle q_{N-1} \vert \hat{V} \vert q_{N-2} \rangle^2 + \langle q_{N-1} \vert hat{V} \vert q_{N-2} \rangle \xi (t_{N-2}) \big ]d\beta - \\ \nonumber
&&\cdots
[\mu \langle q_1 \vert \hat{V} \vert q_0 \rangle^2+ \langle q_1 \vert \hat{V} \vert q_0 \rangle \xi (t_0) \big ]d\beta - \\ \nonumber
&&
[\mu\langle q_0 \vert \hat{V} \vert q' \rangle^2 +\langle q_0 \vert \hat{V} \vert q' \rangle \xi (0) \big ]d\beta \bigg \},
\end{eqnarray}
where
$\hat{\boldsymbol{\xi}}=[\xi(0),\xi(t_0),\cdots,\xi(t_{N-2}),\xi(t_{N-1}),\xi(\beta \hbar)]$,
\begin{equation}
f(\hat{\boldsymbol{\xi}})=\frac{1}{\sqrt{(2\pi)^{N+1} \vert \Sigma \vert }} \exp(-\frac{1}{2} \hat{\boldsymbol{\xi}^{\dagger}} {\Sigma} \hat{\boldsymbol{\xi}}),
\end{equation}
and $\sum_{q_0}\sum_{q_1}\cdots\sum_{q_{N-2}}\sum_{q_{N-1}}$ is the summation over the spin path.
$\langle q_i , q_{i+1} \rangle$ can take four values, $V_{11}$, $V_{22}$, $V_{12}$, and $V_{21}$ when ($\langle q_i \vert $, $\vert q_{i+1} \rangle$) are ($\langle 1 \vert$, $\vert 1 \rangle $), ($\langle 2 \vert$, $\vert 2 \rangle $), ($ \langle 1 \vert$, $\vert 2 \rangle $) and ($\langle 2 \vert$, $\vert 1 \rangle $). So that $\langle q_i \vert \hat{V} \vert q_{i+1} \rangle^2$ can be changed to $\langle q_i \vert \hat{V}\circ \hat{V} \vert q_{i+1} \rangle$.

\bibliographystyle{plain}

\end{document}